\documentstyle[preprint,aps]{revtex}
\begin{document}
\preprint{SSF93-10-01}
\draft
\tighten
\title{Multinucleon mechanisms in ($\gamma$,N) and ($\gamma$,NN) reactions}
\author{Jan Ryckebusch,
Lars Machenil, Marc Vanderhaeghen, \protect{\\} Veerle Van der Sluys
  and  Michel Waroquier}
\address{Laboratory for Theoretical Physics \protect\\Proeftuinstraat 86
\protect\\ B-9000 Gent \protect\\ Belgium}
\date{\today}
\maketitle
\begin{abstract}
The similarities in the experimental indications for multinucleon
mechanisms in $(\gamma,p)$ and $(e,e'p)$ processes are pointed out.
For both types of reactions, the substantial role of
two-nucleon emission processes for transitions to
high excitation energies in the residual nucleus is stressed.  A
microscopic model for the calculation of the two-body knockout
contributions to the inclusive $(\gamma,N)$ reaction is presented.
It is based on an unfactorized formalism for the calculation of
electromagnetically induced two-nucleon emission cross sections. The
model is shown to yield a reasonable description of the
overall behaviour of the
$^{12}$C$(\gamma,p)$ and $^{12}$C$(\gamma,n)$ data at high
excitation energies in the residual nucleus.  In the calculations,
effects from non-resonant and resonant pion exchange currents are
included.  Photoabsorption on
these currents are predicted to produce the major
contributions to the exclusive $^{16}$O$(\gamma,n_0)^{15}$O
 process at photonenergies
above the pion threshold.  Double differential cross sections for photon
induced $pp$ and $pn$ emission from $^{16}$O are calculated and compared
with the data.
\end{abstract}
%\newpage
\section{Introduction}
\label{sec:intro}

The apparent success of the one-body picture in explaining the
quasi-elastic $(e,e')$ results was put
in a different perspective when the longitudinal and transverse
response functions were experimentally separated.
In the one-body picture the virtual photon is assumed to
couple with the
individual nucleons in the target nucleus,
the whole process exhibiting hardly any
medium dependence.
Notwithstanding the
extensive amount of work which has been devoted to a theoretical
understanding of the separated $(e,e')$ data,
a full explanation of both response functions in a consistent model has not
yet been accomplished \cite{Wei89}.

Recent coincidence $(e,e'p)$ measurements have established the significant
role played by multinucleon mechanisms in the quasi-elastic
\cite{Ulm87,Wei90}, dip \cite{Lou86} and
$\Delta_{33}$-excitation domain
\cite{Bag89}.
The existence of multinucleon mechanisms
was evidenced
through a rise in the measured
transverse $(e,e'p)$ response functions at high missing energies
E$_m$=$\omega$-T$_{p}$ in the
(A-1) system \cite{Tak89}.
%At these missing
%energies the exclusive nature of the $(e,e'p)$ process
%can no longer be guaranteed.
This excess strength has been shown to be
unlikely due to rescattering effects, since a comparable rise of the
longitudinal strength would then be expected, an effect which has not
been observed experimentally.  It is worth mentioning that recent
measurements of the separate $(e,e'p)$ structure functions
indicate that a similar
situation seems to occur as for the inclusive $(e,e')$ cross sections :
whereas the one-body picture gives a fair account of the complete
quasi-elastic $(e,e'p)$ cross sections for reactions in which the
residual nucleus is created in a hole state \cite{Die90},
discrepancies turn up when it comes to comparing the separated structure
functions \cite{Spaltro}.
In a recent paper we have illustrated the particular sensitivity of the
longitudinal-transverse $(e,e'p)$
response function to multihadron currents related to
pion-exchange and $\Delta_{33}$ excitation \cite{veerle}.

Whereas the gathered evidence for electron scattering reactions
proceeding in part via multinucleon components is relatively new, over the
years overwhelming evidence for the occurrence of multinucleon components
in real photon reactions has been produced.
Rather than attempting to be complete we mention some illustrative
examples.

\begin{enumerate}
\item[(i)]
Using the tagged photon technique, a Glasgow-Edinburgh-Mainz
collaboration succeeded in measuring the $^{12}$C photoproton cross
section up to high excitation energies in the residual nucleus
\cite{McG86}.  Just as for the $^{12}$C$(e,e'p)$ results of refs.
\cite{Ulm87,Wei90,Lou86,Bag89}, the $^{12}$C$(\gamma,p)$ results of
Ref. \cite{McG86} reported excess strength at high missing energies,
which was shown to be unlikely due to one-body knockout from the
deep-lying hole states.  In the meantime, the findings of Ref.
\cite{McG86} have been confirmed by several independent measurements
\cite{Har88,Luc}.

\item[(ii)]
Strong indications for the occurrence of multinucleon
mechanisms have also been obtained for the exclusive regime.
For a long time it
has been realized that exclusive $(\gamma,n)$ reactions at high photon
energies are good candidates to reveal information about the role of
multihadron currents in photoinduced reactions.  Only one
experiment of this type has been reported up to now.  Exclusive
$^{16}$O$(\gamma,n_0)^{15}$O$(g.s.)$ measurements at MIT-BATES
\cite{Bei89} confirmed the
similarity of $(\gamma,p_0)$ and $(\gamma,n_0)$ angular cross sections
for photon energies ranging from just above the particle emission
threshold, which are typically probing the giant resonance region, to
the $\Delta_{33}$-production region.  This is rather surprising given the
uncharged nature of the neutron and the fact that the squared ratio of
the respective magnetic moments $(\mu _p/\mu
_n)^2$ equals 2.13.

\item[(iii)]
As a third illustrative example of obvious multinucleon
components showing up in photonuclear reactions, we mention the
high-resolution $^{12}$C$(\gamma,p)$ measurements of Refs.
\cite{Luc,Spr90} and the high-resolution
$^{40}$Ca$(\gamma,p)$ measurements of Ref. \cite{Caro}.
  In these experiments a strong feeding of states with
a 2hole-1particle ($2h-1p$)
character has been observed.  In a recent publication
\cite{Ryc92} we have pointed out that the angular cross section for
these transitions can be naturally explained by assuming direct proton
emission following the absorption on the pion-exchange currents. A
similar strong feeding of the $2h-1p$ states has been observed in the
recent $^{12}$C$(e,e'p)$ measurements at high missing momenta at
NIKHEF-Amsterdam \cite{Leon}.
\end{enumerate}

Clearly, photonuclear reactions offer a good testing ground for any
model that aims at describing
the multihadron mechanisms in electromagnetically induced nucleon
emission reactions.  In addition, the similarities in some of the qualitative
features of $(\gamma,p)$ and $(e,e'p)$ are so obvious that a combined
analysis is likely to result in a better insight in
both types of reactions.  This procedure might be
particularly useful to arrive at a better
quantitative understanding of the physics
of the dip and the $\Delta_{33}$ region, for which the $(e,e'p)$ spectra
exhibit similar qualitative features as their $(\gamma,p)$ counterparts
\cite{Bag89,Tak89}.

The multinucleon mechanisms in photonuclear reactions have been
customary interpreted in terms of the phenomenological quasideuteron (QD)
model \cite{Lev51}
in which the photon is assumed to be predominantly absorbed by
$np$ pairs.  This model gives a natural explanation of the measured
excess strength in the $(\gamma,p)$ and $(\gamma,n)$ spectra at high
missing energies.  In the QD phenomenology
the measured nucleon is not exclusive and is
accompanied by an other nucleon which remains either undetected or
gets reabsorbed.  The quasideuteron phenomenology
has been confirmed by double coincidence measurements of the type
$(\gamma,pn)$ \cite{MacG91}.
In line with the predictions of the quasideuteron model,
the $(\gamma,pn)$ data were shown to scale with the missing momentum
${\bf P}={\bf p}_p+{\bf p}_n-{\bf q}_{\gamma}$, the P dependence being
determined by the probability of finding in the target nucleus
a $np$ pair with total momentum $\mid {\bf P} \mid$ and zero separation.

In this paper we present a non-relativistic microscopic model
for the calculation of cross
sections for one and two-nucleon knockout processes
following photoabsorption on
finite nuclei.
Our main focus will be on estimating the effect of two-nucleon emission
processes
to $(\gamma,N)$ cross sections starting from principal grounds.
This involves a microscopic
model for the photoabsorption mechanism and a treatment of the final
state interaction between the escaping nucleons and the residual nucleus.
%Special attention is paid to the structure of the target and residual
%nucleus.
Concerning the final state interaction, we rely on a shell-model approach
to deal with the distortions that the struck nucleons undergo in their
way out of the nucleus.  By doing this we do not have to worry about
spurious contributions to the cross sections due to non-orthogonality
problems.
Within this shell-model framework for the treatment of the
final state interaction
we explore the relevance of pionic and $\Delta_{33}$ degrees of
freedom in inclusive and exclusive photonucleon processes.

The plan of this paper is as
follows. In Sect.~II the formalism is sketched.  This includes
a model for the calculation of $(\gamma,NN)$ cross sections and their
contribution to the $(\gamma,N)$ spectra.  In Sect.~III the numerical
results of the $(\gamma,N)$ and $(\gamma,NN)$ cross sections are
presented.  In particular, Sect IIIA deals with the
contributions stemming from pionic and $\Delta_{33}$
degrees of freedom to exclusive
$^{16}$O$(\gamma,n_0)^{15}$O(g.s.) processes
at higher photon energies.  In Sect. IIIB we
summarize some results of calculations aiming at estimating the
influence of two-nucleon knockout on the $(\gamma,N)$ processes leaving
the residual nucleus in a continuum state above the two-particle
emission threshold.  In Sect.~IIIC the results of the
$^{16}$O$(\gamma,NN)$ calculations are compared with the data.
We conclude with a summary and some outlooks in Sect.~IV.

\section{Formalism}
\label{sec:forma}
In line with the above discussion, photonucleon
spectra reflect multinucleon components in both the discrete (exclusive
$(\gamma,N)$) and the continuous part of their spectrum.  As explained
before there are strong indications that the continuum strength can be
attributed to $(\gamma,pn)$ processes.  Consequently, a model for
two-nucleon emission is essential to calculate $(\gamma,N)$ spectra
above the two-particle emission threshold.

In this Section we will first sketch an unfactorized model for the
calculation of $(\gamma,NN)$ cross sections.  This model does account
for the distortions which the outgoing nucleon pair undergoes in its way
out of the target nucleus and has been described in more detail in Ref.
\cite{Ryc93}.   In the process of calculating the two-nucleon knockout
cross sections, the nuclear structure of the target and the residual
nucleus reflects itself in the two-hole spectral function.  A schematic
model for these spectral functions will be presented. Subsequently, by
integrating the derived $(\gamma,NN)$ cross sections over one of the
nucleon coordinates, we will obtain an expression for the inclusive
$(\gamma,N)$ cross section.  Lastly, we will elaborate upon the
two-body currents on which the initial photoabsorption is assumed to
take place.

%\widetext
\subsection{$(\gamma,NN)$ cross sections and two-hole spectral functions}

In the laboratory frame, the coincidence angular cross section for
a $(\gamma,N_a N_b)$ reaction (Figure \ref{pnsch}) is given by
($\hbar$=c=1) :
\begin{equation}
\frac{d^4 \sigma ^{LAB}} {d \Omega _a d \Omega _b dk_a dk_b} =
 \frac{1} {(2\pi)^5}
\sum_{f} \sum_{m_{s_{a}}m_{s_{b}}} \frac{k_a^2 k_b^2}{2 E_{\gamma}}
\frac{1}{2} \sum_{\lambda} \mid m_F^{fi} \mid ^2 \delta (E_{A-2} + E_{a} +
E_{b} - E_A - E_{\gamma})\;,
\label{cross}
\end{equation}
where the Feynman amplitude $m_F^{fi}$ reads :
\begin{equation}
m_F^{fi}= \left< \Psi_{f}^{(A-2)}(E_x);{\bf k}_a m_{s_{a}};{\bf k}_b m_{s_{b}}
\mid J_{\lambda} ({\bf q}_{\gamma}) \mid \Psi _0 \right> \;.
\label{feynman}
\end{equation}
Here, $f$ is a shorthand notation for all quantum numbers specifying the
eigenstates of the (A-2) system and $E_x$ denotes the positive
excitation energy measured with respect
to the ground state of the residual (A-2)
nucleus.
In the cross section of Eq. (\ref{cross}) we have summed over the
spin projections of the escaping nucleons $(m_{s_{a}},m_{s_{b}})$
and averaged over the initial
photon polarization $\lambda$.  The sum over $f$ extends over
the discrete and the
continuum states of the residual (A-2) nucleus.  We assume
that for the present purposes the target and residual nucleus can be
well described with the aid
of Slater determinants of the independent-particle model.  Further, we
discard all effects due the rescattering (multi-step mechanisms).
%Further, we assume that in case of two-nucleon emission the photon
Within these assumptions the main contribution to the $(\gamma,NN)$
cross sections is supposed to come from direct two-nucleon knockout
following the photoabsorption on a two-body current.  In such a reaction
picture the residual nucleus will be created in a 2 hole $(2h)$ state
relative to the ground state $\mid \Psi _0>$ of the target nucleus :
\begin{eqnarray}
\mid \Psi_{f}^{(A-2)}(E_x) > & \equiv &
\mid (hh')^{-1}J_R M_R > \nonumber \\
& =  & \sum_{m_hm_{h'}} \frac{1}{\sqrt{1+\delta_{hh'}}}
<j_hm_hj_{h'}m_{h'} \mid J_R M_R > \nonumber \\
& & \times  (-1)^{j_h+m_h+j_{h'}+m_{h'}} c_{h-m_{h}}
c_{h'-m_{h'}} \mid \Psi_0 >\;.
\end{eqnarray}

By analogy with the partial-wave expansion techniques which are commonly
employed in a shell-model approach to one-nucleon emission processes and
which are extensively described in Ref. \cite{mahaux}, a
double partial wave expansion has recently been suggested for the
two-nucleon emission case \cite{Ryc93}.  Here, the proper
%proper antisymmetrized
asymptotic behaviour of the
A-body wave function $\mid \Psi _f > \equiv
\mid (hh')^{-1}J_R M_R ;{\bf k}_a m_{s_{a}};{\bf k}_b m_{s_{b}}> $ is
determined by :
\begin{eqnarray}
\left<{\bf r}_1 \mbox{\boldmath$\sigma$}_1, {\bf r}_2
\mbox{\boldmath$\sigma$}_2 \mid \Psi_f
\right> & \stackrel {r_1,r_2 \gg r_A} {\longrightarrow} & \frac {1}
{\sqrt{A(A-1)}} {\cal A}_{2(A-2)} \nonumber \\
 & & \left[ \chi_{\frac{1}{2}m_{s_{a}}} (\mbox{\boldmath$\sigma$}_1)
\left(e^{i{\bf k}_{a} \cdot
{\bf r}_1} + f_{k_{a}}(\theta_a) \frac {e^{ik_ar_1}} {r_1} \right)
\right. \nonumber
\\ & & \times \left.
\chi_{\frac{1}{2}m_{s_{b}}} (\mbox{\boldmath$\sigma$}_2) \left(e^{i{\bf k}_b
{\bf \cdot
r}_2} + f_{k_{b}}(\theta_b) \frac {e^{ik_br_2}} {r_2} \right) \mid
(hh')^{-1} J_R M_R > \right] \;,
\label{Abod}
\end{eqnarray}
and is reached through an expansion in terms of the continuum states
$p(\epsilon l j m)$ of the mean-field potential :
\begin{eqnarray}
\mid \Psi_f > & = & \sum_{lm_ljm}\sum_{l'm_{l'}j'm'} \sum_{JMJ_1M_1}
(4 \pi)^2 i^{l+l'}
\frac {\pi} {2 M_N \sqrt{k_a k_b}}
e^{i(\delta_l+\sigma_l+\delta_{l'}+\sigma_{l'})}
Y_{lm_l}^{*}(\Omega_{k_{a}}) Y_{l'm_{l'}}^{*}(\Omega_{k_{b}})  \nonumber \\
              & & \times <lm_l\frac{1}{2}m_{s_{a}} \mid
jm>  <l'm_{l'}\frac{1}{2}m_{s_{b}}\mid j'm'>
<J_R M_R J_1 M_1 \mid J M > \nonumber \\
& & \times < j m j' m' \mid J_1 M_1 >
\mid (hh')^{-1} J_R ; (p(\epsilon _alj)p'(\epsilon _bl'j')) J_1 ; JM> \;,
\label{wave}
\end{eqnarray}
where $\epsilon _a = k_a^2/2M_a$, $\epsilon _b = k_b^2/2M_b$,
$M_N$ is the nucleon mass, $\delta
_l$ is the central phase shift and $\sigma _l$ the Coulomb phase shift
of the continuum single-particle state $p$.
In Eq. (\ref{Abod}) the operator ${\cal A}_{2(A-2)}$ makes sure that the
total A-body wave function is properly antisymmetrized and $r_A$ is a measure
for the radius of the target nucleus.  Asymptotic wave
functions of the type (\ref{Abod}) refer to a situation in which the
detected nucleon pair has the momentum-spin characteristics
$({\bf k}_a,{\bf k}_b,m_{s_{a}},m_{s_{b}})$ and in which the residual
nucleus is created in the $2h$ state $\mid (hh')^{-1} J_R M_R>$.  The sum
over the partial waves $(l,j,m)$ runs over all continuum states of the
single-particle mean-field potential at a particular excitation energy
$\epsilon$, the latter being set by the kinetic energy of the escaping
nucleon under consideration.  The wave function of Eq. (\ref{wave}) has
been derived under the following normalization convention for the
continuum single-particle states :
\begin{equation}
\varphi_{lj}(r,\epsilon) \stackrel{r \gg r_A} {\longrightarrow}
\sqrt{\frac{2M_N} { \pi k}} \frac {sin(kr-\eta ln(2kr) -
\frac{\pi l}{2} + \delta _l + \sigma _l)} {r}\;.
\end{equation}

In order to calculate the Feynman amplitude of Eq. (\ref{feynman}) it is
convenient to expand the nuclear current operator in terms of its
multipole components.  This is commonly done with the aid of the
electric and magnetic transition operators \cite{For66,Ryc88} :
\begin{equation}
J_{\lambda}(q_{\gamma})= - \sqrt{(2\pi)} \sum_{J \geq 1} i^J \sqrt{2J+1}
(T_{J\lambda}^{el}(q_{\gamma})+\lambda T_{J\lambda}^{mag}(q_{\gamma}))\;.
\label{multi}
\end{equation}
Inserting the Eqs. (\ref{wave}) and (\ref{multi}) in the Feynman
amplitude (\ref{feynman}) and performing some basic manipulations we
obtain the following expression :
\begin{eqnarray}
m_F^{fi} & = & - \sqrt{S_{hh'}(E_x)} \sqrt{2 \pi} \sum _{J \geq 1} i^J \hat{J}
\sum_{lm_ljm}\sum_{l'm_{l'}j'm'} \sum_{J_1M_1}
(4 \pi)^2 (-i)^{l+l'}
\frac {\pi} {2 M_N \sqrt{k_a k_b}} e^{-i(\delta_l+\sigma_l
+\delta_{l'}+\sigma_{l'})}
  \nonumber \\
& & \times Y_{lm_l}(\Omega_{k_{a}}) Y_{l'm_{l'}}(\Omega_{k_{b}})
<lm_l\frac{1}{2}m_s\mid
jm>  <l'm_{l'}\frac{1}{2}m_{s'}\mid j'm'> \nonumber \\
& & \times < j m j' m' \mid J_1 M_1 >
\frac{(-1)^{J_R-M_R+1}}{\hat{J_1}}
<J_R -M_R J \lambda \mid J_1 M_1> \nonumber \\
& & \times [ <p(\epsilon _a l j) p'(\epsilon _b l' j');J_1
\| T^{el}_{J} (q_{\gamma}) + \lambda T^{mag}_{J}
(q_{\gamma}) \|hh' ; J_R > \nonumber \\
&& - (-1)^{j_h+j_{h'}+J_R} <p(\epsilon _a l j) p'(\epsilon _b l' j');J_1
\| T^{el}_{J} (q_{\gamma}) + \lambda T^{mag}_{J}
(q_{\gamma}) \|h'h ; J_R > ]
 \;,
\label{feynb}
\end{eqnarray}
where the function $S_{hh'}(E)$ is equal to the joint
probability of removing two nucleons remaining in the states $h$
and $h'$ from the ground state of the target nucleus and of finding the
resulting system (with (A-2) nucleons)
%in a state characterized by the
%quantum numbers $\mu$
with an excitation energy in the interval (E,E+dE).  The function
$S_{hh'}(E)$ is commonly referred to as the two-hole spectral function
\cite{Kra78} and is defined according to :
\begin{equation}
S_{hh'}(E_x)= \sum_{f} \left|
   \left< \Psi_{f}^{(A-2)}(E_x) \mid c_h c_{h'} \mid
\Psi_0 \right> \right|^2 \;.
\end{equation}

In the actual calculations we have adopted a very
schematic model for the two-hole
spectral functions.  Under the assumption that the removal of a nucleon
does not affect the subsequent removal of a second nucleon, the
function $S_{hh'}$(E) can be approximated by the product of two
probabilities : the probability to remove a nucleon in the state h and
create the (A-1) system at an excitation energy $E'$ and the probability
to remove a nucleon in the state $h'$ from the (A-1) system and create the
(A-2) nucleus at an excitation energy E, or formally :
\begin{equation}
S_{hh'}(E)=\int_{0}^{E}S_h(E')S_{h'}(E-E')dE'\;,
\end{equation}
where the hole spectral function $S_h (E)$ is given by :
\begin{equation}
S_h(E)= \sum_{f} \left| \left< \Psi_{f}^{(A-1)}(E) \mid c_h \mid \Psi_{0}
\right> \right| ^2.
\end{equation}
In the optical model  the hole spectral function is
distributed according to a Breit-Wigner law, centered on the
quasi-particle energy $\mid \epsilon _h$-$\epsilon _F \mid$ (here, $\epsilon
_F$
denotes the Fermi energy)  and with a full width at half
maximum given by 2W(E) \cite{Mah85} :
\begin{equation}
S_h(E)=\frac{1}{\pi}\frac{W(E)}{\left(E-\mid \epsilon _h-\epsilon _F \mid
\right)^2 + \left(
W(E) \right)^2}\;.
\label{Sh}
\end{equation}
Parametrizations for the imaginary part of the optical potential W(E)
are obtained from compilations of experimental data and can e.g. be
found in Ref. \cite{Jeu83}.  In this work we have adopted the
parametrization by Jeukenne and Mahaux \cite{Jeu83} :
\begin{equation}
W(E)=\frac{9E^4}{E^4+(13.27)^4} \;\;\;\; (MeV)\;.
\label{Shpar}
\end{equation}
In Fig. \ref{2hspec} some two-hole spectral functions for $^{16}$O
obtained in the outlined model are shown.  In line with the
results
of the quasi-elastic $^{16}$O$(e,e'p)$ measurements
 regarding the spreading of the hole strength in $^{15}$N \cite{Fru84},
the
quasi-particle energies $\mid \epsilon_h-\epsilon_F \mid$
were determined to be 6~MeV
for the $1p_{3/2}$ and 30~MeV for the $1s_{1/2}$ hole state.
For all results of this paper
the two-hole spectral functions $S_{hh'}$ have been
renormalized to unity : $\int dE S_{hh'}(E)=1$.

\subsection{$(\gamma,N)$ cross sections}
For many years, the $(\gamma,N)$ process for transitions in which the
residual nucleus is created at high excitation energies, has been
interpreted as the QD region with an undetected nucleon of opposite
nature.  Here, we will work out a microscopic model which will put us
in the position to calculate cross sections for these inclusive
processes starting from principal grounds.  In line with the basic
assumption of the QD model, we can assume that an important part of the
$(\gamma,N_a)$ cross section at excitation energies above the
two-particle emission threshold  can be attributed to $(\gamma,N_a N_b)$
processes.  Another mechanism which could be expected to contribute
to $(\gamma,N)$
transitions at high excitation energies in the residual nucleus, is the
exclusive process with excitation of the deep lying hole strength.
Accordingly, we write the $(\gamma,N)$ cross section above the two-particle
emission threshold as the sum of a one-nucleon and a two-nucleon knockout
piece  :
\begin{equation}
\frac{d^2 \sigma ^{LAB}} {d \Omega _a dk_a } =
\left. \frac{d^2 \sigma ^{LAB}} {d \Omega _a dk_a } \right|_{[1]}
+ \left. \frac{d^2 \sigma ^{LAB}} {d \Omega _a dk_a } \right|_{[2]} \;,
\label{split}
\end{equation}
where the two-nucleon piece is determined according to :
\begin{equation}
\left. \frac{d^2 \sigma ^{LAB}} {d \Omega _a dk_a } \right|_{[2]} =
\int d\Omega_b \int_{0}^{\infty} dk_b \frac{d^4 \sigma ^{LAB}}
{d \Omega _a d \Omega _b dk_a dk_b}(\gamma,N_a N_b) \;.
\end{equation}
In the calculation of the two-nucleon knockout contribution,
the $(\gamma,N_a N_b)$ cross section is determined within the model
outlined in the previous subsection.  Since we are working in coordinate
space the integration over the solid angle of the undetected nucleon can
be performed analytically.  After some basic manipulations we find with
the aid of the Eq. (\ref{feynb}) that :
\begin{eqnarray}
& & \left. \frac{d^2 \sigma ^{LAB}} {d \Omega _a dk_a } (\gamma,N_a)
\right|_{[2]}
 =
\sum_{hh'} \int dk_b \int dE_x S_{hh'}(E_x)
\delta (E_{A-2}+E_a+E_b-E_A-E_{\gamma})
\nonumber \\
& & \qquad \qquad \times \sum_{J_R} \sum_{lj} \sum_{l_1' j_1'}
\sum_{l'j'} \sum_{J_1 J_1'}
\sum_{J,J' \geq 1} \sum_{J_2} \frac{A-2}{A}
\frac{k_a E_b}{4M_N^2} \frac{\pi}{E_{\gamma}}
(-i)^{l'-l_1'+J-J'} \widehat{J} \widehat{J'} \widehat{j'} \widehat{j_1'}
\widehat{J_1} \widehat{J_1'} \nonumber \\
& & \qquad \qquad \times P_{J_2}(cos\theta_a)
(-1)^{j-1/2+J_R+j'-j_1'} e^{-i(\delta_{l'}+\sigma_{l'}-\delta_{l_1'}
-\sigma_{l_1'})} \frac{1}{2} \left[ 1+ (-1)^{l'+l_1'+J_2} \right]
\nonumber \\
& & \qquad \qquad \times
<J' \; -1 \; J \; 1 \mid J_2 \; 0> <j'\;1/2\;j_1'\;-1/2 \mid J_2 0 >
\left\{ \begin{array}{lll}
    j_1' \; J_1' \; j  \nonumber \\
    J_1 \; j' \; J_2
    \end{array} \right\}
\left\{ \begin{array}{lll}
    J_1 \; J \; J_R  \nonumber \\
    J' \; J_1' \; J_2
    \end{array} \right\} \nonumber \\
& & \qquad \qquad  \times \left\{ \left( 1+(-1)^{J+J'+J_2} \right) \left[
 {\cal M}_{pp';hh'}^{el}(J_1,J,J_R)
\left( {\cal M}_{pp_1';hh'}^{el}(J_1',J',J_R)\right) ^* \nonumber
\right. \right. \\
& & \qquad \qquad  \left.  \qquad \qquad \qquad \qquad \qquad \qquad
+ {\cal M}_{pp';hh'}^{mag}(J_1,J,J_R)
\left( {\cal M}_{pp_1';hh'}^{mag}(J_1',J',J_R)\right) ^* \right] \nonumber \\
& & \qquad \qquad  +\left( 1+(-1)^{J+J'+J_2+1} \right) \left[
 {\cal M}_{pp';hh'}^{el}(J_1,J,J_R)
\left( {\cal M}_{pp_1';hh'}^{mag}(J_1',J',J_R)\right) ^* \nonumber
\right. \\
& & \qquad \qquad
\left. \left. \qquad \qquad \qquad \qquad \qquad \qquad
+ {\cal M}_{pp';hh'}^{mag}(J_1,J,J_R)
\left( {\cal M}_{pp_1';hh'}^{el}(J_1',J',J_R)\right) ^* \right] \right\} \;,
\label{gncross}
\end{eqnarray}
where $\widehat{J} \equiv \sqrt{2J+1}$, the $P_J$ are the familiar
Legendre Polynomials of degree J, $E_x$ is the excitation energy
in the (A-2) nucleus and the two-body matrix elements
${\cal M}$ have been defined according to :
\begin{eqnarray}
{\cal M}_{pp';hh'}^{el,mag}(J_1,J,J_R) & = &
<p(\epsilon_blj) p'(\epsilon_al'j');J_1
\| T^{el,mag}_{J} (q_{\gamma}) \|hh' ; J_R > \nonumber \\
&& - (-1)^{j_h+j_{h'}+J_R} <p(\epsilon_b l j) p' (\epsilon_a l' j');J_1
\| T^{el,mag}_{J} (q_{\gamma}) \|h'h ; J_R >\;,
\end{eqnarray}
where $\epsilon_a^2=k_a^2/(2M_N)$.

The one-nucleon knockout contribution to the cross
sections (\ref{split}) is calculated with a coupled-channel continuum
RPA technique.  For an elaborate description of this model the
interested reader is referred to Ref. \cite{Ryc88}.  In brief, the RPA
model involves a coupled-channel calculation for all one-nucleon
emission channels ($(\gamma,p)$ and $(\gamma,n)$) leaving the residual
nucleus in a hole state relative to the ground state of the target
nucleus.

An obvious shortcoming of the standard RPA is that it does not
account for the spreading of the single-particle hole strength in the
residual nucleus.  Essentially, in the calculation of the cross sections
for a particular reaction channel $C$ (in the RPA formalism a channel $C$
is characterized by the
quantum numbers of the hole state excited in the residual nucleus and
the momentum of the outgoing nucleon $C(n_h l_h j_h ; k_a \frac{1}{2}
m_{s_{a}})$),
it is assumed that all hole strength is concentrated in the residual
nucleus at an excitation energy $\mid \epsilon_h -\epsilon_F \mid$.
Here,
$\epsilon_h$ is the Hartree-Fock single-particle energy of the considered
state.
In order to account for the spreading of the deep-lying hole strength in
the residual nucleus, we have folded the calculated $(\gamma,N)$
angular cross sections with excitation of particular hole state $h$
(denoted by $d \sigma ^{LAB} / d \Omega_a \mid _{RPA(h)}$),
with the hole spectral
function $S_h$ as defined in the preceeding subsection.   The single-nucleon
knockout contribution to the Eq. (\ref{split}) is then given by :
\begin{equation}
\left. \frac{d^2 \sigma ^{LAB}} {d \Omega _a dk_a }
(\gamma,N_a) \right|_{[1]}
= \sum_{h} S_h (E_x) \left. \frac {d \sigma^{LAB}} {d \Omega _a}
\right|_{RPA(h)}
\end{equation}
where the sum extends over all occupied single-particle states in the
residual nucleus and the excitation energy in the residual nucleus $E_x$
is determined by $E_x + S_p = E_{\gamma} + k_a^2/2M+T_{A-1}$.

\subsection{Absorption mechanisms}
\label{sec:absor}
The next step is to provide a model for the dominant mechanisms in the
photoabsorption process.
Since our main focus will be on photon energies below 200~MeV,
we assume the photon to couple predominantly with
the pion dominated nucleon-nucleon correlations in the target nucleus.
These correlations include terms with and without an intermediate
$\Delta_{33}$ excitation, as indicated in Fig. \ref{absor}.  For the
terms with no $\Delta_{33}$ lines we have considered the currents
associated with the one-pion exchange potential (OPEP)
\begin{equation}
V_{\pi}({\bf k})=- \frac{f_{\pi NN}^2}{m_{\pi}^2}
\frac{1}{m_{\pi}^2+{\bf k}^2}
(\bbox{\sigma}_1 \cdot {\bf k}) \; (\bbox{\sigma}_2 \cdot {\bf k} )
\; \bbox{\tau} _1 \cdot \bbox{\tau} _2 \;,
\label{OPEP}
\end{equation}
where $m_{\pi}$ is the pion mass and $f_{\pi NN}$ the pseudovector
pion-nucleon coupling constant, $f_{\pi NN}^2/(4\pi)=0.079$.  The one-pion
exchange current originating from the coupling of an external
electromagnetic field with two nucleons interacting through the
potential (\ref{OPEP}) can be found in many textbooks and is a sum of
the seagull (diagram (a)) and the pion-in-flight term (diagram (b))
\cite{towner}.

In the evaluation of the $\Delta_{33}$ propagators in the diagrams (c)
and (d) we
have introduced an energy-dependent $\Delta_{33}$ decay width
$\Gamma_{\Delta}$,
such that the propagators read
\begin{equation}
\frac{1}{M_{\Delta}-M_N-E_{\gamma}-\frac{i}{2} \Gamma_{\Delta}(E_{\gamma})} \;,
\end{equation}
with $M_{\Delta}$=1232~MeV.  The $\Delta$-decay width $\Gamma_{\Delta}$ is
considered to be exclusively the result of $\Delta \rightarrow \pi + N$
decay and has been determined according to the expression given  in
 Ref. \cite{Oset} :
\begin{equation}
\Gamma_{\Delta}(E_{\gamma}) \approx \frac{8f_{\pi NN}^2}{12 \pi}
\frac{\left(E_{\gamma}^2-m_{\pi}^2 \right)^{3/2}} {m_{\pi}^2}
\frac{(M_{\Delta}-M_N)} {E_{\gamma}} \;.
\end{equation}
In coordinate space, the $\Delta_{33}$-isobar
current corresponding to the diagrams (c) and (d) is then :
\begin{eqnarray}
{\bf J}^{(\pi \bigtriangleup)}({\bf r},{\bf r}_1,{\bf r}_2) & = &
\frac {2 f_{\gamma N \bigtriangleup} f_{\pi N \bigtriangleup} f_{\pi
NN}} {9 m_{\pi}^3 (E_{\Delta} - E_{\gamma}-{i \over 2}
\Gamma_{\Delta}(E_{\gamma}) )}
\left\{ \left[ \left(\mbox{\boldmath$\tau$}_1 {\bf \times}
\mbox{\boldmath$\tau$}_2
\right)_z \mbox{\boldmath$\sigma$}_2 {\bf \cdot} \mbox{\boldmath$\nabla$}_2
(\mbox{\boldmath$\sigma$}_1 \times
\mbox{\boldmath$\nabla$}_2) \times (\mbox{\boldmath$\nabla$}_1 +
\mbox{\boldmath$\nabla$}_2) \right. \right. \nonumber \\ +
& & \left. \left . 4 (\mbox{\boldmath$\tau$}_2)_z \mbox{\boldmath$\sigma$}_2
{\bf \cdot} \mbox{\boldmath$\nabla$}_2
(\mbox{\boldmath$\nabla$}_1 {\bf \times} \mbox{\boldmath$\nabla$}_2)
 \delta({\bf r} - {\bf r}_1)
 \right]
+ 1 \longleftrightarrow 2 \right\}\frac {e^{-m_{\pi}
\mid {\bf r}_2 - {\bf r}_1 \mid}}
      {4 \pi \mid {\bf r}_2 - {\bf r}_1 \mid}\;,
\label{eq:pidelta}
\end{eqnarray}
with $E_{\Delta} \equiv M_{\Delta}-M_N$.
In the absence of a convincing microscopic theory, we are forced to
treat the $\pi NN$ vertex in a phenomenological way.  In this paper we
shall resort to the widely used  monopole form~:
\begin{equation}
F(\Lambda_{\pi},{\bf p})=\frac{\Lambda_{\pi}^2-m_{\pi}^2}
{\Lambda_{\pi}^2+{\bf p}^2} \;,
\end{equation}
with a cutoff parameter $\Lambda _{\pi}$=1.2~GeV, a value obtained in
investigations in which
nucleon-nucleon scattering data are fitted in
terms of the Bonn one-boson exchange potential.

\section{Results}
\label{sec:resul}
All results presented below have been obtained with single-particle wave
functions obtained from a Hartree-Fock (HF) calculation with an effective
interaction of the Skyrme type~:~SkE2 \cite{War87}.  Apart from the
bound state wave functions, the HF calculation determines the distorting
potential in which the partial waves and phase shifts for the escaping
nucleons
are calculated.  All results presented below were checked not to depend
dramatically on this particular choice for generating the mean-field
characteristics of the target nucleus.  This observation can be
understood by considering that the two-body matrix elements are not very
sensitive to the high-momentum components in the mean-field wave functions.

\subsection{Calculations for
the exclusive $^{16}$O$(\gamma$,n$_{0})^{15}$O(g.s.) reaction}
Figure \ref{gn0} contains the calculated exclusive
$^{16}$O$(\gamma,n_{0})^{15}$N cross sections
at three different values of the photon
energy, all lying above the pion threshold.  The results are obtained in
a direct knockout reaction formalism including
the diagrams as listed in Fig. \ref{diahole}.
Given the uncharged nature of the
neutron, the one-body component is restricted to the magnetization
current.
The two-body components involve the currents of Sect. \ref{sec:absor}.
Inspecting Fig. \ref{gn0}
it is clear that the calculations meet our expectations in the sense
that the one-body mechanism represents but a fraction of the measured
strength and that the angular cross sections are determined by neutron
emission following the absorption on the two-body currents.   At forward
neutron angles a destructive interference effect between the one-body
and the MEC contribution is noticed.  Consequently, the cross section is
dominated by the resonant $\Delta_{33}$ term at forward neutron angles.
Remark further that the $\Delta_{33}$ contribution gains in importance with
increasing photon energies.  Whereas at E$_{\gamma}$=150 and 200 MeV a
fair description of the data can still be obtained with the one-body and the
nonresonant MEC contribution, the $\Delta_{33}$ produces the major contribution
at E$_{\gamma}$=250 MeV.  All curves drawn in Fig.~\ref{gn0} are
obtained for a spectroscopic factor of 0.5.  This corresponds with the
ground state in $^{15}$O exhausting 50~\% of the total $1p_{1/2}$ hole
strength.

\subsection{($\gamma$,p) and ($\gamma$,n) reactions above the two-particle
emission threshold}
The results of our model calculations for the $^{12}$C$(\gamma,n)$
reaction above the $pn$ emission threshold
are summarized in Fig. \ref{abogn}.  In all figures of this subsection,
the missing energy is
defined according to $E_m = E_{\gamma}-T_N$, with $T_N$ the kinetic energy
of the detected nucleon in the LAB system.  In Fig.~\ref{abogn}
we have plotted the calculated
contributions from both one- and two-nucleon knockout. The contribution
from exclusive one-neutron knockout (at the considered missing energies
dominated by $1s$-shell removal) is
calculated in the RPA formalism as outlined in Ref. \cite{Ryc88}.  In
previous papers, it was illustrated that the RPA gives a fairly
realistic account of the exclusive $(\gamma,N)$ data below 100 MeV
photon energy \cite{Ryc92,Ryc88}.
For the spreading of the single-hole strength in the
residual nucleus we used the hole spectral function of
Eq. (\ref{Sh}) in combination with the
parametrization of Eq. (\ref{Shpar}).  In line with the $^{12}$C$(p,2p)$
results \cite{Jac73}
which find the s-shell knockout strength being distributed in the
form of a wide peak between 10 and 30 MeV excitation energy, the
quasi-particle energy $\mid \epsilon_h - \epsilon_F \mid$ was fixed at
20~MeV.  The neutron separation energy $S_n$ being 18.7~MeV,
 the peak of the
$s$-shell removal strength corresponds
with a missing energy of about 39 MeV.
Accordingly, the s-shell knockout strength is concentrated just
above the $pn$ threshold (S$_{pn}$=27.4~MeV)
in the $(\gamma,n)$ spectrum.  It should be
stressed that the missing-energy dependence of the $1s$ strength is
mainly determined by the hole spectral function $S_h$.  Whereas, at
E$_{\gamma}$=75~MeV the
$1s$ $(\gamma,N)$ strength is still substantial in comparison with the
$pn$ strength, it is hardly visible at E$_{\gamma}$=100~MeV.  Generally, for
E$_{\gamma} \geq$ 100~MeV the RPA predicts the one-body knockout
contributions from the deep-lying hole states to be a negligible
fraction of the measured $(\gamma,N)$ strength in the continuum.
  The calculated $(\gamma,pn)$
contributions are predicted to be
substantially larger but are observed not to fully exhaust the measured
strength.  Nevertheless the calculations seem to account for the overall
missing
energy behaviour of the data.  This is particularly the case when the
$(\gamma,pn)$ contribution is arbitrarily  renormalized with a factor of two.

The results of the $^{12}$C($\gamma,p)$ calculations are shown in Figs.
\ref{abogp1}, \ref{abogp2} and \ref{emdepen} and compared with
unpublished results of a
%under the kinematical conditions of the
Glasgow-Mainz collaboration \cite{Har293}.  The calculations have been
performed at two photonenergies, one at each side of the pion production
threshold~: $E_{\gamma}=123$ and 150~MeV.  The missing energy
dependencies at different values of the proton angle are drawn in Figs.
\ref{abogp1} and \ref{abogp2}, whereas in Fig. \ref{emdepen} the
full angular
cross sections are shown for different values of the
proton energy.  The proton kinetic energies were chosen such that they
span the whole missing-energy spectrum. From Figs. \ref{abogp1} and
\ref{abogp2}
it emerges that the $(\gamma,pn)$ calculations give a reasonable
description of the photoproton spectra, particularly for the backward
proton angles.  Nevertheless, in
line with the $^{12}$C$(\gamma,n)$ findings presented earlier, the
$pn$ calculations tend to
underestimate the measured photonucleon cross sections.
In particular, this seems to
be the case at forward proton angles and large missing energies.
  The excess strength at higher missing energies, which
corresponds with slow detected protons,
is likely to be due to other mechanisms, like three-nucleon emission
(S$_{ppn}$=34.0~MeV, S$_{pnn}$=35.8~MeV) which are totally discarded in
our calculations.  A striking feature of the
data is the considerable amount of experimental strength which
is observed for the
forward proton angles in the region of the
$pn$ threshold.  Right at the threshold this strength is unlikely to be
due to
two-nucleon knockout.
Furthermore, exclusive proton removal from the $1s_{1/2}$
shell as calculated in the RPA, was found to represent but
a very small fraction of the measured strength for the two considered
photon energies.

Regarding the missing-energy behaviour, similar
features as for the $^{12}$C$(\gamma,p)$ are found in $^{40}$Ca.  The
missing energy behaviour of the $^{40}$Ca$(\gamma,p)$ cross sections for
a fixed value of $E_{\gamma}$ is presented in Fig. \ref{carol}.
Obviously, the predicted $(\gamma,pn)$ strength does not account for the
experimental strength at forward proton angles, whereas a better
description is reached at backward angles.  Remark further the
considerable amount of measured photoproton strength in the region  of
the $pn$ threshold at $\theta_p = 60^{\circ}$.
In the process of calculating the contribution of exclusive
one-nucleon knockout to the $^{40}$Ca$(\gamma,p)$
spectrum, we considered removal from the $1s$ and $1p$ shell in
addition to the 1d$_{5/2}$ orbit.

The effect of the final state interaction of the struck nucleons with
the residual nucleus has been estimated by doing the
$^{12}$C$(\gamma,p)$ calculations at $E_{\gamma}$=150~MeV,
with a plane wave description for the outgoing nucleon pair and comparing
the results with the full distorted-wave cross sections.  In the
formalism outlined in Sect.~II, a plane wave description can be simply
achieved
by replacing the partial waves $p(\epsilon l j m)$ by properly
normalized spherical Bessel functions.   In passing it is worth
mentioning that the plane wave (PW) description does no longer guarantee the
orthogonality between the inital and final states, such that spurious
contributions could enter the cross sections.  From Fig. \ref{emdepen}
it becomes obvious that the effect of the
distortions on the angular cross sections is not too dramatic, but for
the region just above the threshold (T$_p$=100~MeV).  This
particular behaviour can be easily
explained by considering that from energy-conservation arguments
a fast moving proton will be necessarily accompanied
by a slow neutron, and therefore one of the outgoing nucleons
will be subject to strong interactions with the
residual system.

In Fig. \ref{emdepen} the effect of different absorption mechanisms is
also studied.  From the curves drawn in Fig. \ref{emdepen} it is clear
that even at photonenergies as low as 150~MeV
a considerable fraction of the $pn$ strength at backward proton angles
is related to the resonant terms in the nuclear current.  In passing it
is worth mentioning that strictly speaking also the $pp$ channel could
be expected to
contribute to the $(\gamma,p)$ spectra.  Within our model assumptions,
the $pp$ channel can only be fed through the $\Delta_{33}$ diagrams of
Fig. \ref{absor}.   The other diagrams are closed for two-proton
emission since they involve a charge-exchange mechanism.  In line with
the experimental observations \cite{Dor93}, however, the calculated $pp$
strength
represents but a small fraction of the photoabsorption strength emerging
in the $pn$ channel \cite{Ryc93}.  As will become clear in the fortcoming
subsection,
this finding even holds in the region of the $\Delta_{33}$ resonance.

\subsection{Results of the ($\gamma$,pp) and ($\gamma$,pn) calculations}
\label{sec:pppn}
{}From the results presented in previous subsection, it emerged that there
are strong indications that at high excitation energies in the residual
nucleus the measured $(\gamma,N)$ cross sections should not be
interpreted as the result of an exclusive process but reflect
substantial two-nucleon knockout contributions.
In this sense, the $(\gamma,N)$
spectra above the two-nucleon emission threshold, could be expected to
be largely set by the physics of $(\gamma,N_a N_b)$ processes.  The latter
type of reactions, however, offer some supplementary degrees of freedom
which might be worth exploiting in order to reach a better understanding
of two-nucleon mechanisms in finite nuclei.  At present,
a full determination of the
fivefold differential cross sections $d^4 \sigma / d \Omega _a dk_a
d \Omega _b dk_b$ is clearly at the edge of experimental
feasibility.  Recently, however, several labs have produced double
coincidence data at fixed kinematical conditions for one of the outgoing
nucleons \cite{Dor93,Are91}.
To calculate the measured cross sections, we can employ
Eq. (\ref{gncross}). This expression was derived by integrating the
full coincidence $(\gamma,N_a N_b)$
cross section over one of the outgoing nucleon coordinates
and produced the predictions for the two-nucleon components in the
$(\gamma,N)$ spectra.  In order to get some idea regarding the realistic
character of the proposed $(\gamma,NN)$ model, it is worth checking its
predictions against the data.  Here, we present some calculations under the
kinematical conditions of the measurements of Ref. \cite{Are91}.

The results of the $^{16}$O$(\gamma,pn)$ and $^{16}$O$(\gamma,pp)$
calculations at $E_{\gamma}$=281~MeV and different values of the proton
energy are summarized in Figs. \ref{gpn1} and \ref{gpn2}.  At this
photonenergy the cross sections are dominated by the $\Delta_{33}$
current.  The $(\gamma,pn)$ and $(\gamma,pp)$ cross
sections are found to exhibit similar characteristics.  For the high
kinetic energies, the angular cross sections are clearly forwardly
peaked.  With decreasing proton kinetic energy flatter distributions are
obtained. The data compromise the proton energy dependence of the cross
section at a fixed proton angle ($\theta _p=52^{\circ}$).
The comparison with the data is shown in Fig. \ref{gpn3}.
At low proton energies, the
calculated $(\gamma,pp)$ and $(\gamma,pn)$ clearly underestimate the
data.  This region is usually interpreted as being dominated by pion
production. Our calculations seem to suggest that even at low proton
kinetic
energies there is a considerable background of direct two-nucleon emission.
Apart from the pion knockout,
also three and more nucleon knockout processes will preferentially feed
the low proton energy domain of the spectrum.  For
lack of a microscopic
theory for three and more nucleon ejection processes, explicit
$(\gamma,p \pi)$ measurements will be needed to gain insight into
the pionproduction channels.
At higher kinetic energies, the calculations give a reasonable account
of the $pp$ and $pn$ emission channel.  The dashed curve for the
$(\gamma,pp)$ channel is the cross section obtained with a plane-wave
description for the escaping protons.  It is clear that the
distortion effects from the FSI reduce the peak of the cross section and are
substantial in explaining the data.  Remark further how the background
of $pp$ strength at low proton energies can be partly ascribed
to FSI effects.  The $pp$ cross sections are
about one order of magnitude smaller than the $pn$ cross sections, a
feature which is nicely reproduced by the calculations.

\section{Conclusion}
Summarizing, we have presented a microscopic study of multinucleon
mechanisms in photoinduced nucleon-knockout processes.
Our study encompasses both inclusive and exclusive photonucleon
processes in addition to two-nucleon knockout reactions.

In the exclusive regime, we have reported on $^{16}$O$(\gamma,n_0)$
results above the pion-production threshold.  Here, the predominant role
of pionic degrees of freedom, including the $\Delta_{33}$ excitation, in
photonucleon processes is striking.

Our main focus has been on the contribution from two-nucleon knockout to
the inclusive photonucleon spectra.
The description
relies
on an unfactorized approach to two-nucleon knockout reactions.  The fair
description of the $^{16}$O$(\gamma,pn)$ and $^{16}$O$(\gamma,pp)$ data,
makes us feel rather confident about
the realistic character of the employed two-nucleon knockout
formalism.
Regarding the $(\gamma,N)$ processes, we have
shown that at higher missing energies
the $pn$ emission strength largely
exceeds the strength related to one-nucleon
removal from the deep-lying hole states.  We find our model, which
accounts for photoabsorption on the resonant and non-resonant pion
currents, to give a reasonable description of the general features of
the $(\gamma,N)$ spectra at
high excitation energies in the residual nucleus.  Nevertheless, the
calculations tend to
selectively underestimate the available $^{12}$C$(\gamma,p)$ and
$^{12}$C$(\gamma,n)$ spectra above the $pn$ threshold.  This is
particularly the case at forward nucleon angles and higher
missing energies.  This feature, together with the observation that
quite some strength resides in the region of the $pn$ threshold, points
towards other mechanisms, besides $pn$ emission,
contributing to the photonucleon processes
with excitation of the residual nucleus in a continuum state.
It would be worth investigating this in more detail, particularly in
view of the fact that recent calculations predict the short-range
effects to occur mainly at high excitation energies \cite{ciofi,mutter}.

Finally, we stress that the techniques adopted in this
paper can be easily applied to $(e,e'N)$ and $(e,e'NN)$ processes.
It is to be hoped
that a combined analysis of the $(\gamma,p)$ and $(e,e'p)$ spectra,
together with new data from $(\gamma,NN)$ and $(e,e'NN)$
measurements, will
lead to a better insight into the nature  of the multinucleon mechanisms
in electromagnetically induced nucleon knockout.  Given their particular
sensitivity to multinucleon mechanisms, reactions with real photons will
play a substantial role in this program.

\acknowledgements
One of us (J.R.) is indebted to D. van Neck for useful discussions
on two-hole spectral functions.
The authors would like to thank P.~Harty and C. Van den Abeele
 for giving us the permission to
show their data prior to publication.  We are also indebted to
K.~Heyde for valuable discussions and a careful reading of the manuscript.
This work was supported by the National Fund for Scientific
Research and in part by the NATO through the research grant NATO-CRG920171.

\begin{figure}
\caption{Kinematics for the $(\gamma,N_a N_b)$ reaction.   The figure
sketches the situation in which the photon and the escaping nucleons
remain in one plane (planar kinematics).}
\label{pnsch}
\end{figure}

\begin{figure}
\caption{Two-hole spectral functions for $^{16}$O as calculated with the
schematic model outlined in the text.  The two-hole spectral functions
have been normalized to unity.}
\label{2hspec}
\end{figure}

\begin{figure}
\caption{Feynman diagrams included in the evaluation of the two-body
matrix elements. \qquad \qquad}
\label{absor}
\end{figure}

\begin{figure}
\caption{Diagrammatic representation of an exclusive $(\gamma,N)$ process
with one- and two-body absorption mechanisms in a direct knockout picture.}
\label{diahole}
\end{figure}

\begin{figure}
\caption{Calculated
$^{16}$O($\gamma$,n$_{0}$)$^{15}$O(g.s., (1p$_{1/2}$)$^{-1}$) angular cross
sections in a direct knockout model at three values of the photon
energy.  Dotted line : photoabsorption on the magnetization current.
Dashed line : photoabsorption on the magnetization and pion-exchange current.
Solid line :
photoabsorption on the magnetization, pion-exchange
and $\Delta_{33}$-isobar
current.  The
data are from Ref. \protect{\cite{Bei89}}.}
\label{gn0}
\end{figure}

\begin{figure}
\caption{Missing energy dependence of the $^{12}$C($\gamma,n$) cross
section at $\theta_n$=66$^{\circ}$.  The dotted line shows the
calculated cross sections for one-body knockout from the 1s$_{1/2}$
shell. The dashed line represent the contribution from $(\gamma,pn)$.
The solid line gives the sum of both contributions.  For the dot-dashed
line the $(\gamma,pn)$ contribution has been arbitrarily multiplied with
a factor of two.  The data are from Ref. \protect{\cite{Har88}}.}
\label{abogn}
\end{figure}

\begin{figure}
\caption{Missing energy dependence of the $^{12}$C$(\gamma,p)$ cross
section at E$_{\gamma}$=123~MeV.  The solid line shows the prediction
of the $(\gamma,pn)$ calculations. The data are from Ref.
\protect{\cite{Har93}}.}
\label{abogp1}
\end{figure}

\begin{figure}
\caption{As in Fig. \protect{\ref{abogp1}}
but at E$_{\gamma}$=150~MeV. \qquad \qquad \qquad \qquad \qquad \qquad
\qquad \qquad \qquad}
\label{abogp2}
\end{figure}

\begin{figure}
\caption{Angular $^{12}$C$(\gamma,p)$
cross sections at different values of the missing energy for
$E_{\gamma}$=150~MeV.  The solid line is the prediction of a
$(\gamma,pn)$ calculation with all MEC and $\Delta_{33}$ diagrams of
Fig. \protect{\ref{absor}}.  For the dotted line only the MEC
diagrams are accounted for.  The dot-dashed line is the equivalent of
the solid curve but is obtained with a plane wave description for the
outgoing nucleons.}
\label{emdepen}
\end{figure}

\begin{figure}
\caption{Missing energy bahaviour of the $^{40}$Ca$(\gamma,p)$ cross section
at $E_{\gamma}$=60~MeV. The data are from Ref. \protect{\cite{caro93}}.
The dot-dashed shows the calculated contribution from one-proton
knockout. The dashed line represents the contribution from $(\gamma,pn)$.
The solid line gives the sum of both contributions.}
\label{carol}
\end{figure}

\begin{figure}
\caption{$^{16}$O$(\gamma,pn)$ angular cross sections at
$E_{\gamma}$=281~MeV and several values of the proton kinetic energy.}
\label{gpn1}
\end{figure}

\begin{figure}
\caption{As in Fig. \protect{\ref{gpn1}} but for the $^{16}$O$(\gamma,pp)$
process. \qquad \qquad \qquad \qquad \qquad \qquad
\qquad \qquad \qquad}
\label{gpn2}
\end{figure}

\begin{figure}
\caption{Proton energy dependence of the $^{16}$O$(\gamma,p)$,
$^{16}$O$(\gamma,pn)$ and $^{16}$O$(\gamma,pp)$ reaction
 at E$_{\gamma}$=281~MeV and
$\theta_p$=52$^{\circ}$.  In the upper figure ($^{16}$O$(\gamma,p)$)
both the calculated contribution from $^{16}$O$(\gamma,pp)$ (dotted
line),
$^{16}$O$(\gamma,pn)$ (dot-dashed line) and their sum (solid line) are shown.
For the $(\gamma,pp)$ channel the dashed curve gives the plane-wave result.
The data are from Ref. \protect{\cite{Are91}}.}
\label{gpn3}
\end{figure}

\end{document}